\documentclass[conference,onecolumn]{IEEEtran}
\ifCLASSINFOpdf
  \usepackage[pdftex]{graphicx}
\else
   \usepackage[dvips]{graphicx}
\fi

\usepackage{amsmath}
\usepackage{amsfonts}
\usepackage{amsmath,epsfig,subfigure}
\usepackage{latexsym,amssymb}
\usepackage{cite,url}
\usepackage{lettrine}
\usepackage{hyperref}
\def\draftmode{}
\ifx\draftmode\undefined
\newcommand{\comment}[1]{}
\else
\newcommand{\comment}[1]{ \marginpar{$\Longleftarrow$}{\bf $<$#1$>$} }
\fi

\begin{document}

\title{Towards Dynamic Real-Time Geo-location Databases for TV White Spaces} 

\author{\IEEEauthorblockN{Ahmed Saeed and Mohamed Ibrahim}
\IEEEauthorblockA{Dept. of Comp. Sc. and Eng.\\
Egypt-Japan University for Sci. and Tech.\\
Alexandria, Egypt\\
\{ahmed.saeed,mohamed.ahmed\}@ejust.edu.eg}
\and \IEEEauthorblockN{Moustafa Youssef}
\IEEEauthorblockA{Wireless Research Center\\
Egypt-Japan Univ. of Sc. and Tech. and\\
 Alexandria University, Alexandria, Egypt\\
Email: moustafa.youssef@ejust.edu.eg}
\and \IEEEauthorblockN{Khaled A. Harras}
\IEEEauthorblockA{Computer Science Department\\
School of Computer Science\\
Carnegie Mellon University
 \\
kharras@cs.cmu.edu }}

\maketitle

\begin{abstract}
Recent FCC regulations on TV white spaces allow geo-location
databases to be the sole source of spectrum information for White Space
Devices (WSDs). Geo-location databases protect TV band incumbents by
keeping track of TV transmitters and their protected service areas based on
their location, transmission parameters and sophisticated propagation models.
In this article, we argue that keeping track of both TV transmitters and
TV receivers (i.e. TV sets) can achieve significant improvement in
the availability of white spaces. We first identify wasted spectrum opportunities, both temporal and spatial, due to the current approach of white spaces detection. We then propose \emph{DynaWhite}, a cloud-based architecture that orchestrates the detection and dissemination of highly-dynamic, real-time, and fine-grained TV white space information. \emph{DynaWhite} introduces the next generation of geo-location databases by combining traditional sensing techniques with a novel unconventional sensing approach based on the detection of the passive TV receivers using standard cell phones. We present a quantitative evaluation of the potential gains in white space availability for large scale deployments
of \emph{DynaWhite}. We finally identify challenges that need to be addressed in the research community in order to exploit this potential for leveraging dynamic real-time fine-grained TV white spaces.

\end{abstract}

\IEEEpeerreviewmaketitle

\section{Introduction}

The unlicensed usage of TV white spaces, which refer to portions of the UHF spectrum
(and parts of the VHF spectrum in the US), has been introduced
by the FCC as a means to supply mobile devices'
ever increasing demand for high quality communication and
multimedia streaming \cite{press2}. Utilizing these
white spaces is only allowed while
strictly forbidding interference with primary spectrum incumbents
(i.e. TV receivers and wireless microphones). The ruling ensures
the mitigation of interference between spectrum incumbents
and White Space Devices (WSDs) through forcing WSDs to use
one of two methods: Following the first method, WSDs use
white spaces after sensing the spectrum for TV transmissions
with a very low threshold of $-107$ dbm.
Spectrum sensing capabilities add complexity and cost
complications to WSDs, especially with this
low threshold. The second method, which is the currently preferred one, relies on consulting
geo-location databases that keep track of available
white spaces in certain areas \cite{press2}. Although the main spectrum
incumbents that need protection are the TV \textbf{\emph{receivers}}, not
transmitters, TV receivers are typically \textbf{\emph{passive}}, i.e. they
do not transmit signals, and thus are difficult to detect\footnote{Protection
of wireless microphones is guaranteed in the FCC regulation by providing two channels specifically for wireless microphones and allowing events with large numbers of wireless microphones to register in the geo-location databases.}. Therefore, geo-location databases protect TV receivers by protecting the entire coverage area of a TV transmitter through keeping record of TV transmitters' information including location, antenna height,
transmission power, and channels used. The geo-location databases combine this information with sophisticated propagation models in order to determine the protection area of a TV transmitter, where no WSD can be active \cite{gurney2008geo,Senseless}.

In this article, we argue that current geo-location databases
regulations, while guaranteeing high protection
of the spectrum incumbents, waste significant spectrum opportunities
by protecting the entire coverage area of TV \textbf{\emph{transmitters}}.
In particular, it is not necessary that the entire coverage area
contains active, i.e. turned on, TV \textbf{\emph{receivers}}, which are the main devices to
be protected. In many cases, there are spectrum holes, both temporal
and spatial, that are void of active TV receivers. In particular, recent studies show that, while
average Americans watch 5.2 hours of TV a day, less than 10\% of these TV viewers watch broadcast channels \cite{press3}. These temporal spectrum holes are wasted opportunities that can reach up
to 23 channels in some urban areas (e.g. Miami City, Florida) according to our results presented in this article.
This potential gain in spectrum availability is a great incentive for leveraging dynamic real-time TV white spaces awareness, especially in spectrum-hungry urban areas that will experience exponential demand on wireless bandwidth.

Based on the aforementioned observations on current geo-location databases,
we propose \emph{DynaWhite}; a new cloud-based architecture for future dynamic real-time TV white spaces spectrum awareness.
 \emph{DynaWhite} approach leverages both the TV transmitters and receivers information to provide fine-grained real-time spectrum availability information; For \emph{TV transmitters} information, \emph{DynaWhite} extends the current geo-location databases \cite{gurney2008geo,Senseless}, that use over-protective propagation models, to allow for collaborative spectrum sensing using WSDs \cite{akyildiz2011cooperative}. Although such solutions are not attractive in terms of cost, accuracy, and power consumption; it is expected that sensing will take an important role in the future of WSDs as spectrum becomes more congested and technology advances.     
 For \textbf{\emph{TV receivers}} information, to handle the problem of sensing the passive TV receivers, \emph{DynaWhite} uses an \textbf{unconventional sensing approach} that leverages the ubiquitous standard cell phones and other mobile devices.
 In particular, we argue for a crowd-sourcing approach, where today's sensor-rich cell phones can be used to detect TV receivers' location and state (e.g. ON/OFF and TV channel viewed) based on their acoustic, visual, and other fingerprints\footnote{Different sensors, e.g. the microphone and camera, can be used to detect the presence of the TV set as well as the current running channel as we show in \cite{dynawhitereport}.}.

Finally, \emph{DynaWhite} uses the cloud computing scalability and vast computational and storage capabilities to keep track of this highly-dynamic, real-time and fine-grained geo-location TV white spaces.

This article is organized as follows. Section II identifies the wasted white spaces opportunities according to the description of current geo-location databases by the FCC \cite{press2}. We then present \emph{DynaWhite}, our next generation highly-dynamic, real-time and fine-grained TV white space detection architecture in Section III. We evaluate the potential gain that would be achieved with the adoption of \emph{DynaWhite} in Section IV. Section V provides a discussion of the research challenges that need to be addressed for truly realizing this architecture. We finally present related work and conclude the article in sections VI and VII.

\section{TV White Spaces: The Wasted Opportunities}
\label{sec:waste}

Initially, FCC regulations mandated WSDs to use both geo-location databases
and spectrum sensing. Later, a recent modification of the FCC regulations
allowed WSDs to obtain information on available white spaces \emph{only}
through geo-location databases \cite{press2}. While, this new regulation
lessens the burden on WSDs, it suffers from wasting white spaces both
spatially and temporally.

\begin{figure*} [!t]
\centering
      \includegraphics[width=0.7\textwidth]{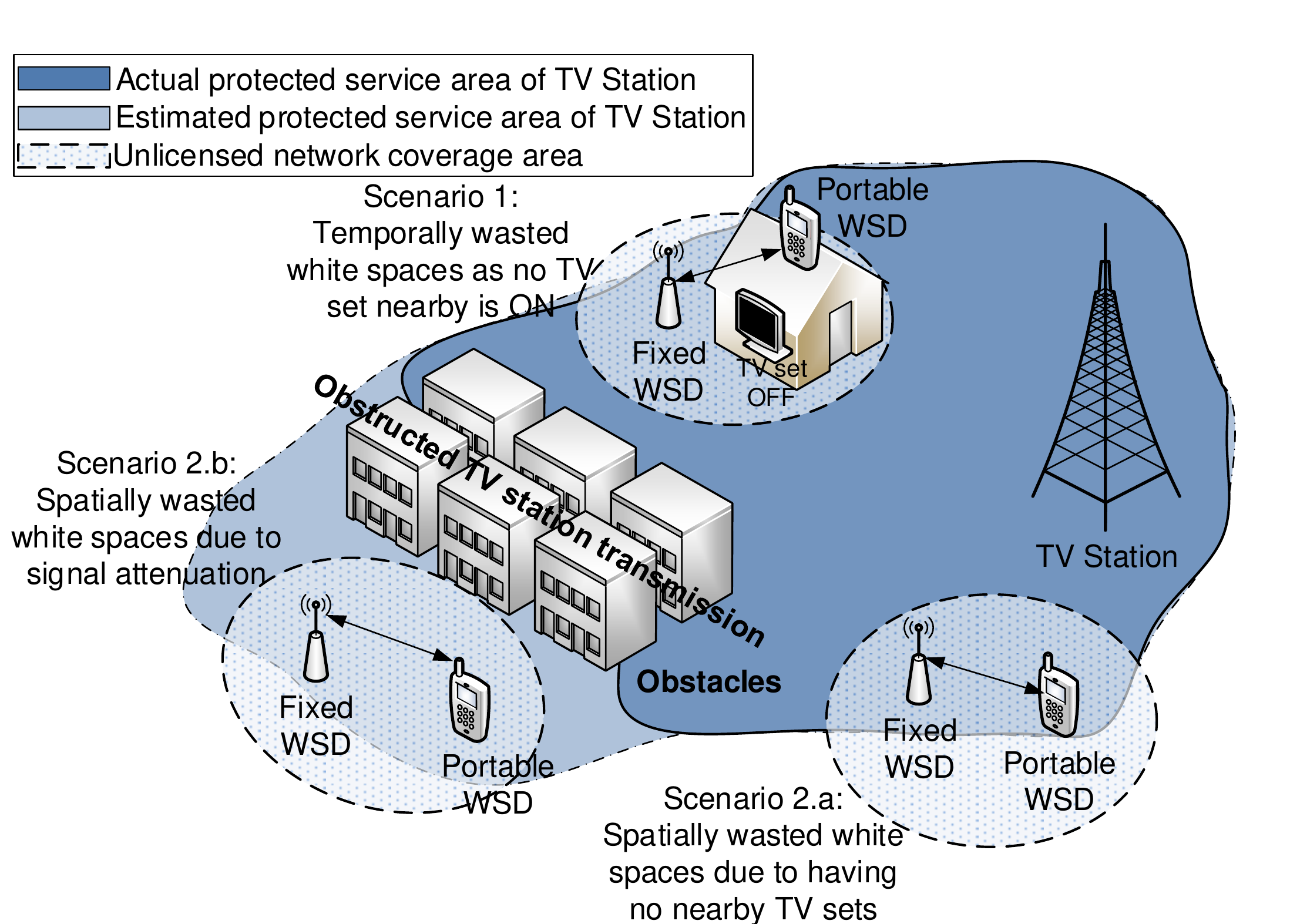}
  \caption{Three different scenarios in which white spaces opportunities were lost.}
  \label{fig:spec_opp}
\end{figure*}

Figure~\ref{fig:spec_opp} summarizes two different scenarios
in which white space opportunities are missed. In Scenario 1, the temporally wasted opportunity, although the white space network lies
within the TV station's protected service area, the TV set in the vicinity
of the network is turned OFF (or for the same practical purpose, currently
tuned to a particular channel, leaving other channels available for
unlicensed usage). Conventional geo-location databases, that are not
aware of the location of TV sets or their state, will declare the
entire set of allocated channels in the area as unavailable.

On the other hand, the second scenario is composed of
two different instances (a and b) in which white spaces are spatially wasted. In Scenario 2.a, the white space network lies within
the protected service area but has no TV sets in the vicinity
of the network. In Scenario 2.b, the white space network
lies in the theoretical protected service area of the TV station,
but the station's signal is obstructed. These scenarios lead to wasted white space opportunities, because conventional geo-location databases rely solely on propagation models.

Avoiding wasting spectrum opportunities is particularly important
in urban areas, where huge numbers of wireless devices and
congested RF spectrum are the norm. Table~\ref{table1} shows
the available white spaces for both fixed and portable WSDs
in four urban and four rural cities using the ShowMyWhiteSpace
application developed by the FCC approved geo-location database company Spectrum Bridge Inc. \cite{ShowWhiteSpaces}.
We observe that white spaces are more available for portable WSDs
compared to fixed WSDs as the maximum allowable transmission
power for them is 4W and 40mW respectively\cite{press2}.

\begin{table}[!t]
\centering
\begin{tabular}{||p{1in}||p{1.2in}||p{1.5in}|p{1.5in}||}
\hline
\hline
\multicolumn{4}{||c||}{Urban Areas}\\
\hline
City& Available Channels for Fixed WSD (4W)& Available Channels for Portable 100 mW WSD& Available Channels for Portable 40 mW WSD\\
\hline
\hline New York, NY & 0 & 0&0\\
\hline Los Angeles, CA& 0 &0& 0\\
\hline  Miami, FL& 0  & 0&2\\
\hline Philadelphia, PA & 1 &0 &3\\

\hline
\hline
\hline
\multicolumn{4}{||c||}{Rural Areas}\\
\hline \hline Hudson, NY& 13 & 7 &18\\
\hline Palatka, FL &    15 & 11 &20 \\
\hline Amador City, CA &    18 & 10    &20 \\
\hline Conconully, WA &    23 & 16    &25 \\
\hline
\hline
\end{tabular}
\caption
{Urban and rural areas channel availability using ShowMyWhiteSpace application by Spectrum Bridge Inc (as of Feb 11$^{th}$, 2012). \cite{ShowWhiteSpaces}.}
\label{table1}
\end{table}

To sum up, temporal spectrum opportunity waste occurs when the state of the TV set (ON/OFF or the channel it is currently tuned to) is ignored. However, spatial spectrum opportunity waste occurs when an area with no TV sets, or outside the coverage area of a TV tower, is falsely protected, in order to over protect TV receivers. These wasted opportunities are critical especially in spectrum-hungry urban areas.
\emph{DynaWhite}'s novel unconventional TV receivers sensing approach combined with collaborative spectrum sensing using WSDs address these wasted opportunities.

\section{The DynaWhite Architecture}
\label{sec:dynawhite}

The identified wasted spectrum opportunities require alterations in geo-location databases to evolve from storing and processing the relatively static TV towers' information (e.g. location, transmission power, antenna height and channel), to collect, process and store the highly dynamic, real-time TV sets' information. In order to address this evolution, we exploit the scalability and reliability of cloud storage and processing. We propose the \emph{DynaWhite} architecture that can generate real-time geo-location databases via a cloud-based system that is responsible for processing and aggregating sensory information. This information is different in: (1) nature, e.g. TV set state information, and (2) reliability metrics, e.g. based on the quality of the contributing sensors. This architecture aims to address the expected bandwidth demand on white space networks by tracking every available spectrum opportunity in real-time. For the rest of this section,
we present an overview of the proposed architecture and its operation, then we discuss
the two main components of the new database: the conventional WSD spectrum
sensing for TV transmitters and the \emph{unconventional} TV receiver state sensing using standard cell phones.

\subsection{Overview and Operation}

\begin{figure*} [!t]
\centering
      \includegraphics[width=0.58\textwidth]{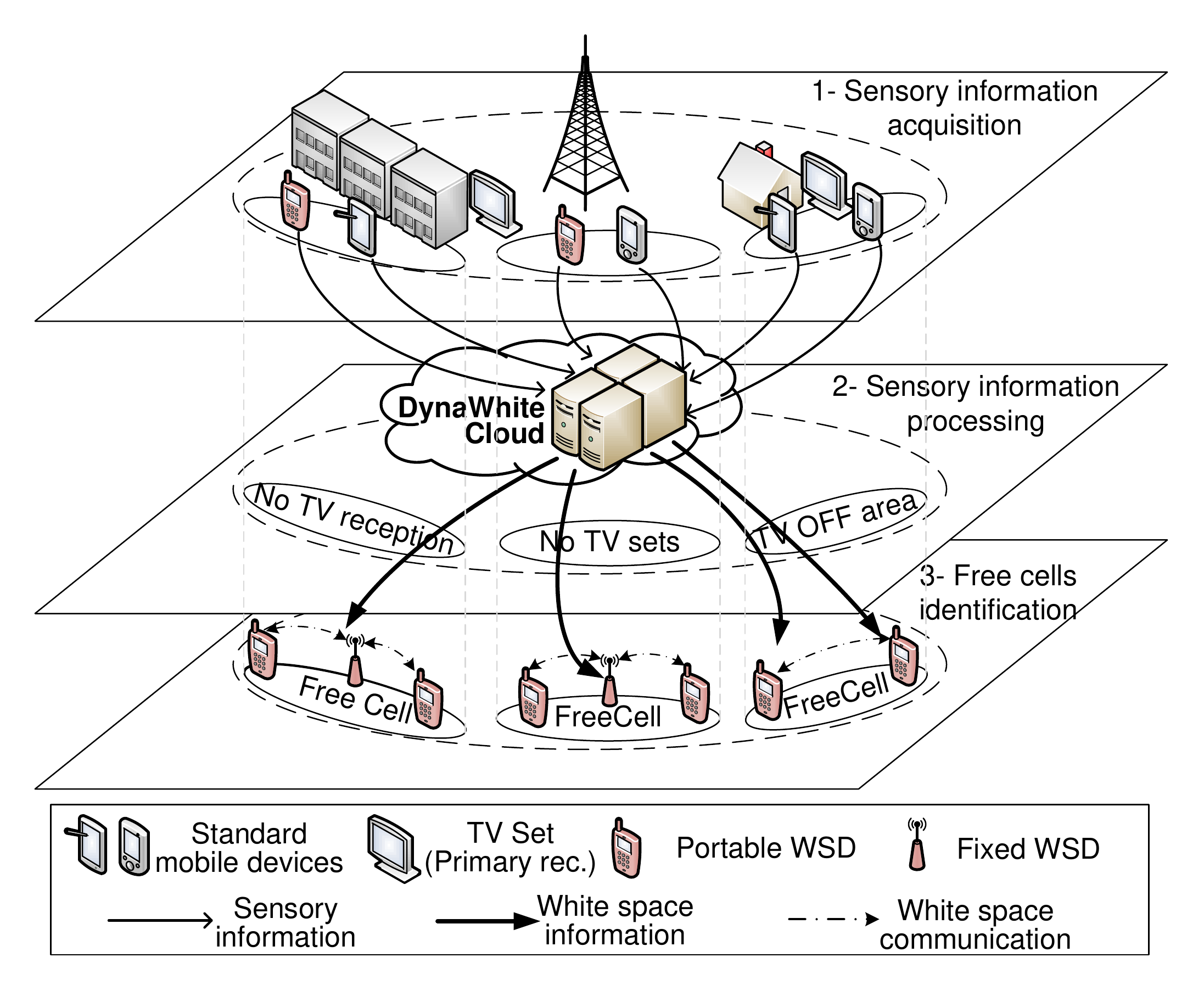}
  \caption{DynaWhite operation scenario.}
  \label{fig:arch}
\end{figure*}

Figure \ref{fig:arch} gives an overview of a scenario of \emph{DynaWhite}.
Contributing devices, equipped with localization
mechanisms, submit location-tagged spectrum sensing data (from WSDs) or ambience sensing data (from standard cell phones) to the \emph{DynaWhite} cloud infrastructure. \emph{DynaWhite} divides the area of interest into fine grained cells and calculates white space availability for each cell separately based on the collected sensory information. A cell is considered a
``free cell'' with respect to a certain channel in two cases: (1) when there is no TV signal on the channel within the cell or (2) when the cell has no TV sets that are tuned to the channel. The architecture supports that devices submit their data (push mode) or to query devices at specific locations for spectrum status on demand (pull mode). The two modes enable the system to enhance its view of the spectrum. It should be noted that, \emph{DynaWhite} defaults to traditional geo-location databases when there is no sensory information available for a particular cell or when the confidence of the spectrum information is not high.

\begin{figure} [!t]
\centering
      \includegraphics[width=0.8\textwidth]{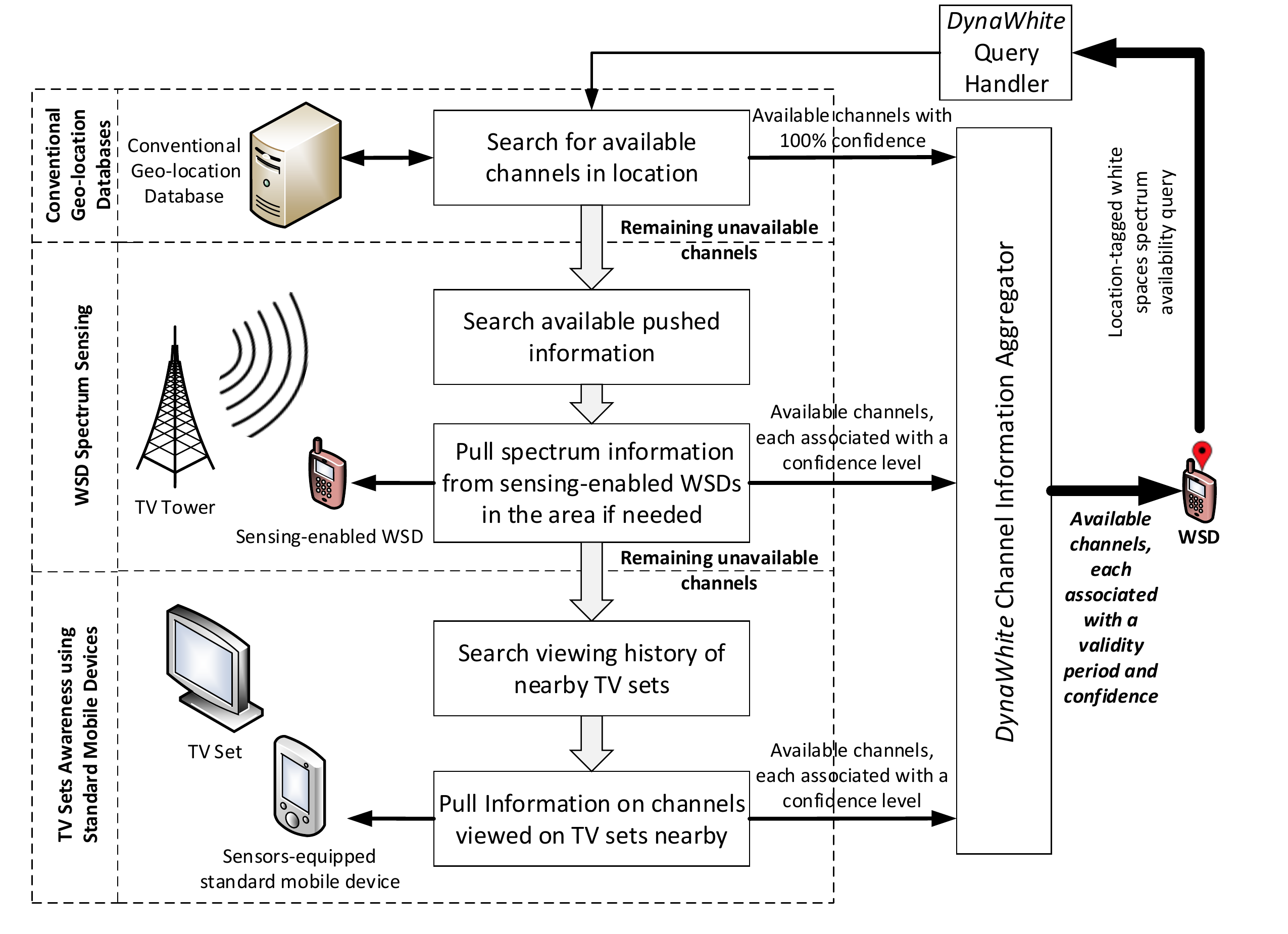}
  \caption{DynaWhite query processing flow.}
  \label{fig:layered_arch}
\end{figure}

Figure~\ref{fig:layered_arch} shows how a typical spectrum availability query is handled by \emph{DynaWhite}. The multi-layered flow is divided based on the different types of available white space opportunities. These opportunities vary in terms of persistence and reliability. For each location-annotated \textit{White Spaces Information Query}, \emph{DynaWhite} aims to obtain the most persistent and reliable availability information. We define white spaces information \emph{persistence} as the duration of time the spectrum information will be valid. Moreover, we define white spaces information \emph{reliability} as a function of \emph{DynaWhite}'s confidence in the contributors and the conclusion made based on their collected sensory information.

The first layer consults conventional geo-location databases to identify white space opportunities based only on the parameters of licensed TV \emph{transmitters}. This layer presents white space information with the highest reliability and persistence.

The second layer checks the presence of TV transmitters based on spectrum sensing readings previously collected by WSDs that lie within the white space network's coverage area. If there is no data available, \emph{DynaWhite} will pull real-time spectrum information on demand. This layer can increase the number of available white space channels or help the network enhance its transmission parameters based on the sensed interference on each channel. The white space information obtained from this layer is persistent as the view of the spectrum is not likely to change fast. However, this sensory information needs to be confirmed by more than one contributing device in order to establish its reliability.

The third layer uses standard cell phones to detect the presence and the state of TV sets within the white space network's coverage area. This is particularly useful when the TV transmitter is operational while no TV sets are tuned to its channel. This type of white space information is highly volatile because TV viewers switch channels randomly. Moreover, in order to increase the reliability in such information more than one contributor needs to support the same sensor based decision.
We note, however, that the information collected over time from multiple devices at a particular location
can determine the set of channels typically viewed at this location,
ultimately increasing the persistence and reliability of the spectrum availability data by basing
its decisions on the history of channels viewing. 

White space opportunities, obtained from each layer, are associated with a validity period to improve \emph{DynaWhite}'s reliability. The WSD is required to re-check with the
\emph{DynaWhite} database for the white space's availability after the validity period expires. This period depends on both the persistence level and reliability associated with each reading. These levels are calculated based on the number of contributors conforming on the same reading as well as the source of the sensed information. Upon the reception of \emph{DynaWhite}'s response to its query, the WSD can determine the channels to transmit on, based on its required quality of service and the validly period of each available white space.

\subsection{WSD Spectrum Sensing}
\label{sec:sensing}

Signal readings are collected from spectrum sensing enabled WSDs.
Spectrum sensing enabled devices as specified by the FCC are either sensing only devices, that depend solely on spectrum sensing to detect available channels, or as the FCC encourages, mixed devices that use spectrum sensing with geo-location databases \cite{press2}. Another source of spectrum information includes dedicated sensing infrastructures, that could be deployed in order to enhance spectrum utilization in spectrum starved areas. Several cooperative spectrum sensing algorithms can be used to make the best of these spectrum sensory information \cite{akyildiz2011cooperative}.

FCC regulations for sensing only devices require initial sensing of 30 seconds to select the channel to be used. After that, these devices are required to ensure that the channel is vacant every 60 seconds \cite{press2}. While these regulations ensure protection of TV spectrum incumbents, they require a mobile device to regularly check to ensure the consistency of the spectrum view. On the other hand, \emph{DynaWhite}'s interest lies in determining the spectrum view for each cell and the cell's area is smaller than the approved sensing only device coverage area. Therefore, the spectrum sensing requirements for \emph{DynaWhite} are more relaxed.

\subsection{TV Set Awareness using Standard Mobile Devices}

Sensory information collected by standard mobile devices (cell phones, laptops, tablets, etc) can be used as unconventional spectrum sensors to deduce whether a TV set is available in a certain cell or not. If a TV set is detected, regular checks are performed to check whether the TV is turned ON or not and to detect the channel that TV is currently tuned to. Sensors like camera, microphone, accelerometer, etc can be used to detect both TV set behavior (e.g. visual and acoustic fingerprints) and TV viewer behavior (e.g. sitting, using the remote and texting about TV shows). This information could be used to identify the presence of TV sets and whether they are ON or OFF. Moreover, using online streaming sites and channel guides, the channel currently playing could be detected using only the acoustic fingerprint of the TV set \cite{dynawhitereport}.

Smart infrastructures (e.g. smart homes) can also be used for inferring
information as they are equipped with sensors designated for different
functionalities required to improve the quality of people's life. These functionalities include TV controls which can be directly used to update \emph{DynaWhite}. Another form of sensors are the new generation of Internet-enabled
smart TV sets that can be used to update \emph{DynaWhite}'s database in real-time with information on the channels viewed. These sources of information could reliably tell whether there is a TV that is ON in a certain area of interest and the channel that TV is tuned to.

By accumulating and aggregating detected channel information for each TV
set, collected from different sources with different reliability levels,
and correlating them with time, high confidence TV set detection decisions can be reached. \emph{DynaWhite} then becomes capable of estimating the viewer's watching profile. This stochastic behavior can further be leveraged to enhance \emph{DynaWhite} decisions and reliability estimation.

\section{DynaWhite Potential}
In this section, we quantify the potential gain that can be obtained with the adoption of \emph{DynaWhite} based systems through two city-scale simulations.

\subsection{Potential Gain in White Spaces}

We conducted simulations for Miami City in Florida and New York County in New York, in order to illustrate the potential gains in spectrum availability in urban areas that can be achieved using \emph{DynaWhite}. According to the United State's Census
Bureau, New York County, NY has 732,204 households with an area of $22.83 mi^2$ and Miami City has 149,077 households with an area of $35 mi^2$ \cite{census}. In our simulation, we distributed these households uniformly over the two areas assigning one TV set for each household. We randomly picked only 21\% of the TV sets as operational
 and selected 10\% of the TV sets to be showing a broadcast channel \cite{press3}.
We assigned one of the broadcast TV channels in the designated area to each of the TV sets (27 channels in New York County and 26 channels in Miami City) \cite{ShowWhiteSpaces}.

\begin{table}[!t]

\centering

\begin{tabular}{|l||c|c|c|}
\hline

Power		& Coverage &Min. separation distance for adjacent channel transmission 	&Min. separation distance for co-channel transmission	\\
\hline
\hline 1 mW	& 59 m & 9 m	& 182 m\\
\hline 5 mW	& 86 m & 13.2 m	& 265 m\\
\hline 10 mW	& 101 m & 15.5 m	& 310 m\\
\hline 40 mW 	& 140 m& 22.4 m & 430 m\\
\hline 100 mW	& 173 m& 26.4 m & 533 m\\
\hline

\end{tabular}
\caption
{Parameters used in the simulations for TV band devices with transmission powers 1, 5, 10, 40 and 100 mW in terms of the maximum coverage
distance, the minimum distance between the TV band device, and the TV set to avoid interference \cite{press2}.}
\label{table3}
\end{table}

We distributed 100,000 WSDs over the two areas to measure the potential increase in the number of available channels. Then, we measured the amount of free white spaces on which each WSD can operate without violating the FCC's protection criteria for the TV sets. The protection criteria for co-channel
transmission was selected to be $23 db$ SNR and $-33 db$ SNR for adjacent channel transmissions \cite{press2}. We applied the Okumura-Hata model for urban areas \cite{hata1980empirical} to identify the separation needed between the WSD and the TV set in order to maintain the minimum field strength of $41 dbu$ for TV service at the TV set which is specified by the FCC \cite{press2}. The simulation was made for the following transmission powers:

\begin{itemize}
 \item 1 mW used in experiments conducted for local area white space networks in \cite{bahl2009white}.
 \item 5 and 10 mW account for possible transmission powers that could be used to increase white spaces availability and range.
 \item 40 mW which is the maximum transmission power specified by the FCC for WSDs working
 in within a TV stations protected service area broadcasting on adjacent channel.
 \item and 100 mW which is the maximum allowed transmission power by the FCC for portable WSDs \cite{press2}.
\end{itemize}

The geo-location database's awareness of these different transmission
powers presents another enhancement that enables the detection of white
spaces relative to the WSD's transmission power. This results in avoiding
current geo-location database's assumptions that the WSDs can work only on
two levels of power (i.e. 40mW and 100mW). Table~\ref{table3} summarizes the parameters for each of the different transmission obtained using the Okumura-Hata model for urban areas.

\begin{table}[!t]

\centering
\begin{tabular}{|p{0.5in}|p{1.3in}|p{1.3in}|p{1.3in}|p{1.3in}|}
\hline
Power		& Percent of WSDs gaining more channels in New York County	& Average number of channels gained in New York County & Percent of WSDs gaining more channels in Miami	& Average number of channels gained in Miami  \\
\hline
\hline 1 mW	& 	100\%					& 	9.65 						& 	100\%					& 	23.4\\
\hline 5 mW	& 	99\%					& 	4 						& 	100\%					& 	21\\
\hline 10 mW	& 	92\%					& 	2.7 						& 	100\%					& 	 19.6\\
\hline 40 mW 	& 	49.8\% 					&	1.53 						& 	100\%					& 	 15.5\\
\hline 100 mW	& 	19.9\% 					&	1.27 						& 	100\%					& 	 12.2\\
\hline

\end{tabular}
\caption
{Simulation of potential gain in white spaces for WSDs working in Maimi City, Florida and New York County, New York with
transmission powers 1, 5, 10, 40 and 100 mW given information on TV sets locations and the channels they are currently tuned to.}
\label{table4}
\end{table}

The results of the simulation are summarized in Table~\ref{table4}.
There is a significant difference between the two cities due
to the difference in population densities. Comparing \emph{DynaWhite}'s
spectrum availability to conventional geo-location databases
(Table~\ref{table1}), for Miami City, devices working with any
transmission power obtain at least 12.2 extra channels instead of
no channels in case of 100 mW for conventional geo-location
databases. Similarly, in New York County, despite the high density
of TV sets, almost 20\% of devices working with 100 mW transmission
power obtain 1.27 channels on average to work on instead of not being
able to work at all when using the conventional geo-location databases.

As we lower the transmission power, aiming at increasing the bandwidth
available for enterprise local area networks, the number of available
channels increases reaching 23.4 in Miami and 9.65 in New York County
for all devices working with 1 mW (i.e. having a coverage of 59m). This
highlights the significant gains that can be achieved through a city
scale \emph{DynaWhite} deployment.

\subsection{Research Challenges}
\label{sec:future}

The \emph{DynaWhite} architecture can significantly alter the perception of the white spaces, especially  within urban cities. This section discusses a number of research challenges that need to be tackled by the research community to realize \emph{DynaWhite}'s potential gain.

\subsubsection{Sensory Reliability}

To declare a channel free for unlicensed usage based on crowd-sourced
sensory information, a high confidence level is needed in this information. One of the main challenges faced by \emph{DynaWhite} is to determine the confidence associated with each contributor and contributed sensory information. Another related challenge is to determine the validity period of a free channel.

Measures should also be taken to ensure the reading's validity and to protect \emph{DynaWhite} from malicious users injecting false sensory information to gain more white spaces. Factors determining the validity of the sensory information include the number of the contributors, consistency of different contributions, the confidence associated with each contributor and the history of sensory readings collected for each cell. Viewers' profiles could be used to enhance the reliability in the decisions.

Moreover, the range of validity periods should be studied. For instance, current FCC regulations have a minimum validity period of 60 seconds for portable devices relying on either sensory information or geo-location databases. Lower validity periods should be studied weighing their potential gain as compared to the overhead of accessing \emph{DynaWhite} with high frequency.

\subsubsection{Location Granularity}

Current FCC regulations require a location estimation accuracy
within 50 meters \cite{press2}. The effect of using \emph{DynaWhite},
especially with knowing the location of TV sets, on the accuracy
of localizing WSDs should be studied. On the other hand,
building the dynamic geo-location databases requires the location
information of the TV sets and the sensing devices.
Location estimation can be performed with different granularity.
For example, the location of the building where the TV set is located
can be determined using the GPS sensor on the phone and/or GSM-based
localization. More fine-grained indoor localization systems, e.g. \cite{unlocwang2012no},
can be used to determine the TV sets position to within a couple
of meters. This can enable a new set of applications, e.g.
femto TV white spaces cells. Finally, the granularity of \emph{DynaWhite}'s
view of the map should be determined. In particular, the size
of the cells for which \emph{DynaWhite} detects white spaces availability
needs to be studied.

\subsubsection{Incentives}

Convincing mobile holders and spectrum sensing-enabled device holders to sense and share their sensory data with others to allow for better spectrum utilization is a challenging problem. This is particularly important as users will have to use their scarce battery and network connection to share their information with \emph{DynaWhite} servers. Batch operation can be used to upload the data when, e.g., free WiFi is available or the device is connected to power. However, this will be at the expense of the real-time aspect of the sensed information.

\subsubsection{Privacy}

Another aspect is the privacy of the user who shares her phone sensor information. One solution is to submit just final results of detection, i.e the status of the TV and the current channel instead of sharing the raw data, i.e. acoustic and visual data. However, this may consume the battery faster. Another potential solution is to leverage the user's other more capable less-restricted devices, such as laptops, to do the processing locally.

\subsubsection{Regulations}

The \emph{DynaWhite} system cannot be deployed until the regulatory authorities issue rulings allowing for ambience sensing as a source of spectrum information.
To address this challenge, controlled field-tests should be carried out to
show the feasibility, gains, and any conflicts the proposed fine-grained
geo-location databases can cause.

\section{Related Work}

There are currently two approaches to ensuring the protection of TV white space incumbents both based on TV transmitters information. The first approach, adopted by the FCC, relies on geo-location databases that keep track of TV transmitters' parameters and propagation models in order to estimate the areas that need to be protected
\cite{gurney2008geo}. The work presented in \cite{Senseless} extends this approach through using sophisticated propagation models and presents a scalable architecture of geo-location databases. The second approach, adopted by the IEEE 802.22 standard, relies on collaborative spectrum sensing between the WSDs. In this approach, WSDs submit their spectrum view to a central entity that is responsible for performing spectrum sharing functionalities \cite{akyildiz2011cooperative}. \emph{DynaWhite} incorporates these two approaches and extends them by its unconventional sensing approach for TV receivers.

On the other hand, detecting TV receivers has been addressed before using either special hardware, that senses the power leakage of a receiver's local oscillator \cite{oh2008white, DetectPR}, or using central trusted \emph{manually}-updated databases \cite{gurney2008geo,bezabih2012digital}.
The former technique requires the usage of special hardware that needs to
be setup in the vicinity of the TV set. Such techniques do not scale and are hard to deploy. The latter technique does not scale to a large scale.studies the effect of knowing TV receivers information on the available white spaces, in terms of amount of additional available frequencies, they assume that in some countries, e.g. Norway, everyone that has TV receiver have to register TV receiver information in order to pay the broadcasting license fees.

We believe that \emph{DynaWhite} poses as the next generation of geo-location databases for TV white spaces. This approach aims at detecting all available white space opportunities through a multi-layered architecture. This architecture uses \emph{unconventional} spectrum sensing, supported by conventional spectrum sensing and geo-location databases to detect spectrum opportunities that
are diverse in reliability and persistence. In addition, \emph{DynaWhite} exploits the cloud's computational and storage capabilities to infer available white spaces.

\section{Conclusion}

We presented \emph{DynaWhite}, an architecture for creating and maintaining the next generation geo-location databases characterized by being highly-dynamic, real-time, and fine-grained. The proposed architecture provides new temporal and spatial spectrum opportunities by incorporating both crowd-sourcing based spectrum sensing and an unconventional technique for TV sets localization
and state detection using ambience sensors. Furthermore, we layout possible future directions of research for improving the performance of this new generation of geo-location databases.

\bibliographystyle{IEEEtran}
\bibliography{tv_wmag}

\begin{thebibliography}{10}
\providecommand{\url}[1]{#1}
\csname url@samestyle\endcsname
\providecommand{\newblock}{\relax}
\providecommand{\bibinfo}[2]{#2}
\providecommand{\BIBentrySTDinterwordspacing}{\spaceskip=0pt\relax}
\providecommand{\BIBentryALTinterwordstretchfactor}{4}
\providecommand{\BIBentryALTinterwordspacing}{\spaceskip=\fontdimen2\font plus
\BIBentryALTinterwordstretchfactor\fontdimen3\font minus
  \fontdimen4\font\relax}
\providecommand{\BIBforeignlanguage}[2]{{%
\expandafter\ifx\csname l@#1\endcsname\relax
\typeout{** WARNING: IEEEtran.bst: No hyphenation pattern has been}%
\typeout{** loaded for the language `#1'. Using the pattern for}%
\typeout{** the default language instead.}%
\else
\language=\csname l@#1\endcsname
\fi
#2}}
\providecommand{\BIBdecl}{\relax}
\BIBdecl

\bibitem{press2}
{Federal Register}, ``Unlicensed operation in the {TV} broadcast bands,''
  December 2010.

\bibitem{gurney2008geo}
D.~Gurney, G.~Buchwald, L.~Ecklund, S.~Kuffner, and J.~Grosspietsch,
  ``Geo-location database techniques for incumbent protection in the {TV} white
  space,'' in \emph{New Frontiers in Dynamic Spectrum Access Networks, 2008.
  DySPAN 2008. 3rd IEEE Symposium on}.\hskip 1em plus 0.5em minus 0.4em\relax
  IEEE, 2008, pp. 1--9.

\bibitem{Senseless}
R.~Murty, R.~Chandra, T.~Moscibroda, and P.~Bahl, ``Senseless: A
  database-driven white spaces network,'' \emph{Mobile Computing, IEEE
  Transactions on}, vol.~11, no.~2, pp. 189 --203, feb. 2012.

\bibitem{press3}
{The Nielsen Company}, ``State of the media: Cross-platform report {Q}1 2012,''
  2012.

\bibitem{akyildiz2011cooperative}
I.~Akyildiz, B.~Lo, and R.~Balakrishnan, ``Cooperative spectrum sensing in
  cognitive radio networks: A survey,'' \emph{Physical Communication}, vol.~4,
  no.~1, pp. 40--62, 2011.

\bibitem{dynawhitereport}
A.~Saeed, M.~Ibrahim, K.~Harras, and M.~Youssef, ``Unconventional sensing for
  mobile tv band devices,'' School of Computer Science, Carnegie Mellon
  University, Tech. Rep. CMU-CS-QTR-117, February 2013.

\bibitem{ShowWhiteSpaces}
\BIBentryALTinterwordspacing
{Spectrum Bridge, Inc.}, ``{ShowMyWhiteSpace - Locate TV White Space
  Channels},'' 2010. [Online]. Available:
  \url{http://whitespaces.spectrumbridge.com/Overview/Home.aspx}
\BIBentrySTDinterwordspacing

\bibitem{census}
\BIBentryALTinterwordspacing
{United States Census Bureau}, ``Quick, easy access to facts about people,
  business, and geography.'' [Online]. Available:
  \url{http://quickfacts.census.gov}
\BIBentrySTDinterwordspacing

\bibitem{hata1980empirical}
M.~Hata, ``{Empirical formula for propagation loss in land mobile radio
  services},'' \emph{Vehicular Technology, IEEE Transactions on}, vol.~29,
  no.~3, pp. 317--325, Aug. 1980.

\bibitem{bahl2009white}
P.~Bahl, R.~Chandra, T.~Moscibroda, R.~Murty, and M.~Welsh, ``White space
  networking with {Wi-Fi} like connectivity,'' in \emph{ACM SIGCOMM
  2009}.\hskip 1em plus 0.5em minus 0.4em\relax ACM, 2009, pp. 27--38.

\bibitem{unlocwang2012no}
H.~Wang, S.~Sen, A.~Elgohary, M.~Farid, M.~Youssef, and R.~Choudhury, ``No need
  to war-drive: Unsupervised indoor localization,'' in \emph{Proceedings of the
  10th international conference on Mobile systems, applications, and
  services}.\hskip 1em plus 0.5em minus 0.4em\relax ACM, 2012, pp. 197--210.

\bibitem{oh2008white}
S.~Oh, A.~Naveen, Y.~Zeng, V.~Kumar, T.~Le, K.~Kua, and W.~Zhang, ``White-space
  sensing device for detecting vacant channels in {TV} bands,'' in
  \emph{Cognitive Radio Oriented Wireless Networks and Communications, 2008.
  CrownCom 2008. 3rd International Conference on}.\hskip 1em plus 0.5em minus
  0.4em\relax IEEE, 2008, pp. 1--6.

\bibitem{DetectPR}
B.~Wild and K.~Ramchandran, ``Detecting primary receivers for cognitive radio
  applications,'' in \emph{DySPAN 2005}, 2005, pp. 124 --130.

\bibitem{bezabih2012digital}
H.~Bezabih, B.~Ellingsaeter, J.~Noll, and T.~Maseng, ``Digital broadcasting:
  Increasing the available white space spectrum using {TV} receiver
  information,'' \emph{Vehicular Technology Magazine, IEEE}, vol.~7, no.~1, pp.
  24--30, 2012.

\end{thebibliography}

\end{document}